\documentclass[11pt,a4paper]{article}
\usepackage{times,latexsym}
\usepackage{url}
\usepackage[T1]{fontenc}

\usepackage[acceptedWithA]{tacl2021v1}
\usepackage{xspace,mfirstuc,tabulary}

\newif\iftaclinstructions
\taclinstructionsfalse 
\iftaclinstructions
\renewcommand{\confidential}{}
\renewcommand{\anonsubtext}{(No author info supplied here, for consistency with
TACL-submission anonymization requirements)}
\newcommand{\instr}
\fi

\iftaclpubformat 

\else

\fi

\usepackage{times}
\usepackage{latexsym}
\usepackage{mdframed}
\usepackage{subcaption}

\usepackage[T1]{fontenc}

\usepackage[utf8]{inputenc}

\usepackage{xcolor}

\usepackage[noabbrev,capitalize,nameinlink]{cleveref}
\usepackage{dingbat}
\usepackage{tabularx}
\usepackage{booktabs}
\usepackage{pifont}
\usepackage[dvipsnames]{xcolor}
\usepackage{float}
\usepackage{graphicx}
\usepackage{microtype}
\usepackage{inconsolata}
\usepackage{hyperref}
\usepackage{comment}

\newcommand{\github}{\smash{\raisebox{-0.1cm}{\includegraphics[width=0.4cm]{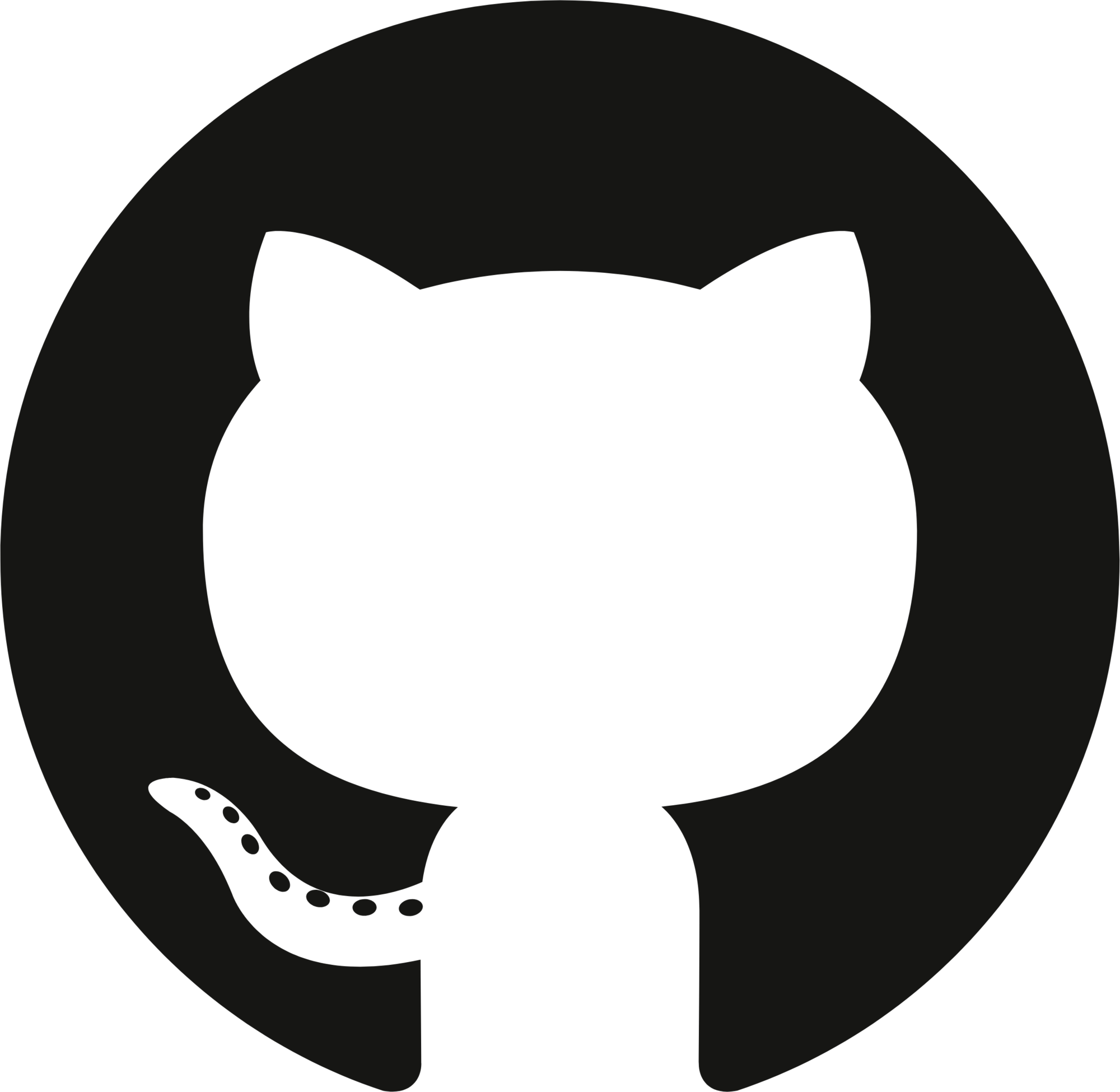}}}}
\newcommand{\anonlink}{https://anonymous.4open.science/r/queer_NLP_bib-5674}

\title{Queer NLP: A Critical Survey on \textcolor{red}{L}\textcolor{black}{iterature} 
\textcolor{orange}{G}\textcolor{black}{aps,} 
\textcolor{yellow!70!black}{B}\textcolor{black}{iases} \textcolor{black}{and} 
\textcolor{green!70!black}{T}\textcolor{black}{rends}}

\author{
    Sabine Weber\textsuperscript{1,2,}\thanks{$^*$First author, please send correspondence to \texttt{sabine.weber@uni-bamberg.de}.} 
    \hspace{.5em}Angelina Wang\textsuperscript{1,3,}\footnotemark[2]
    \hspace{.5em}Ankush Gupta\textsuperscript{1,4,5,}\thanks{$^\dagger$Equal contributions, alphabetical order by first name}
    \hspace{.5em} Arjun Subramonian\textsuperscript{1,}\footnotemark[2] \\
    {\bf
    Dennis Ulmer\textsuperscript{1,6,}\footnotemark[2]
    \hspace{.5em} Eshaan Tanwar\textsuperscript{1,7,}\footnotemark[2]
    \hspace{.5em} Geetanjali Aich\textsuperscript{1,8,}\footnotemark[2]
    \hspace{.5em} Hannah Devinney\textsuperscript{1,9,}\footnotemark[2]}\\
    {\bf
    Jacob Hobbs\textsuperscript{1,17,}\footnotemark[2]
    \hspace{.5em} Jennifer Mickel\textsuperscript{1,10,}\footnotemark[2]
    \hspace{.5em} Joshua Tint\textsuperscript{1,11,}\footnotemark[2]
    \hspace{.5em} Mae Sosto\textsuperscript{1,12,}\footnotemark[2]}\\
    {\bf Ray Groshan\textsuperscript{1,13,}\footnotemark[2]
    \hspace{.5em} Simone Astarita\textsuperscript{1,}\footnotemark[2]
    \hspace{.5em}  Vagrant Gautam\textsuperscript{1,14,}\footnotemark[2] 
    \hspace{.5em} Verena Blaschke\textsuperscript{1,15,16,}\footnotemark[2]}\\
    {\bf
    William Agnew\textsuperscript{1,17,}\footnotemark[2]
    \hspace{.5em}  Wilson Y Lee\textsuperscript{1,18,}\footnotemark[2]
    \hspace{.5em} Yanan Long\textsuperscript{1,19,}\footnotemark[2]}\\
    \small
    \textsuperscript{1}Queer in AI
    \hspace{.5em}\textsuperscript{2}University of Bamberg
    \hspace{.5em}\textsuperscript{3}Cornell Tech
    \hspace{.5em}\textsuperscript{4}IIIT-DELHI
    \hspace{.5em}\textsuperscript{5}PayGlocal\\
    \small
    \hspace{.5em}\textsuperscript{6}ILLC, University of Amsterdam
    \hspace{.5em}\textsuperscript{7}University of Copenhagen
    \hspace{.5em}\textsuperscript{8}University of Massachusetts Amherst \\
    \small
    \hspace{.5em}\textsuperscript{9}Linköping University
    \hspace{.5em}\textsuperscript{10}EleutherAI 
    \hspace{.5em}\textsuperscript{11}Arizona State University\\
    \small
    \hspace{.5em}\textsuperscript{12}Centrum Wiskunde \& Informatica (CWI)
    \hspace{.5em}\textsuperscript{13}University of Maryland, Baltimore County\\
    \small
    \hspace{.5em}\textsuperscript{14}Heidelberg Institute for Theoretical Studies (HITS)
    \hspace{.5em}\textsuperscript{15}LMU Munich
    \hspace{.5em}\textsuperscript{16}MCML\\
    \small
    \hspace{.5em}\textsuperscript{17}Carnegie Mellon University
    \hspace{.5em}\textsuperscript{18}HubSpot
    \hspace{.5em}\textsuperscript{19}StickFlux Labs\\
}

\date{\today}

\begin{document}
\maketitle
\begin{abstract}
Natural language processing (NLP) technologies are rapidly reshaping how language is created, processed, and interpreted by humans. With current and potential applications in hiring, law, healthcare, and other areas that impact people's lives, understanding and mitigating harms towards marginalized groups is critical. In this survey, we examine NLP research papers that explicitly address the relationship between LGBTQIA+ communities and NLP technologies. We systematically review all such papers published in the ACL Anthology up until February 2026 (n=122), to answer the following research questions: (1) What are current research trends? (2) What gaps exist in terms of topics and methods? (3) What areas are open for future work? We find that while the number of papers on queer NLP has grown within the last few years, most papers take a reactive rather than a proactive approach, focusing on shortcomings of existing systems rather than creating new solutions. Our survey uncovers many opportunities for future work, especially regarding stakeholder involvement, intersectionality, interdisciplinarity, and languages other than English. We also offer an outlook from a queer studies perspective, highlighting understudied topics and blind spots in the harms addressed in NLP papers. Beyond being a roadmap of what has been done, this survey is a call to action for work towards more just and inclusive NLP technologies.

\github\hspace{0.15cm}\href{https://github.com/webersab/queer_NLP_bib}{\footnotesize \texttt{webersab/queer\_NLP\_bib}}
\end{abstract}

\section{Introduction}

In this survey, we examine the state of NLP research as it relates to LGBTQIA+ topics, which we call Queer NLP (for an in-depth definition, see Section \ref{sec:def}). We use the term \textit{queer} as a shorthand for the broad spectrum of identities represented under the LGBTQIA+ umbrella (see \hyperref[sec:glossary]{Glossary}, \S\ref{sec:glossary}), while recognizing the diverse and heterogeneous experiences within this community. Although there has been a significant amount of work in NLP on bias more broadly~\cite{stanczak2021surveygenderbiasnatural, field-etal-2021-survey, goldfarb-tarrant-etal-2023-prompt}, we foreground the queer community by surveying papers that discuss the distinct social impacts queer people face in the context of NLP technologies.

As evidenced by an increasing number of articles and its first archival event for ACL (Queer in AI @ NAACL 2025; \citealp{queerinai-ws-2025-main}), Queer NLP is growing as a topic of study, which necessitates synthesizing and systematizing the knowledge in this subfield to understand current trends, identify gaps, and assess promising future directions.
These objectives are critical for NLP to meaningfully engage with queer communities and their linguistic realities to foster more inclusive and socially responsible systems.
Therefore, we systematically survey all Queer NLP papers from the ACL Anthology. We annotate these papers for tasks, types of data, queer groups and harms they focus on, as well as their motivations, methods, and engagement with intersectionality and stakeholders. 

\begin{figure}[t]
    \centering
    \includegraphics[width=0.985\linewidth]{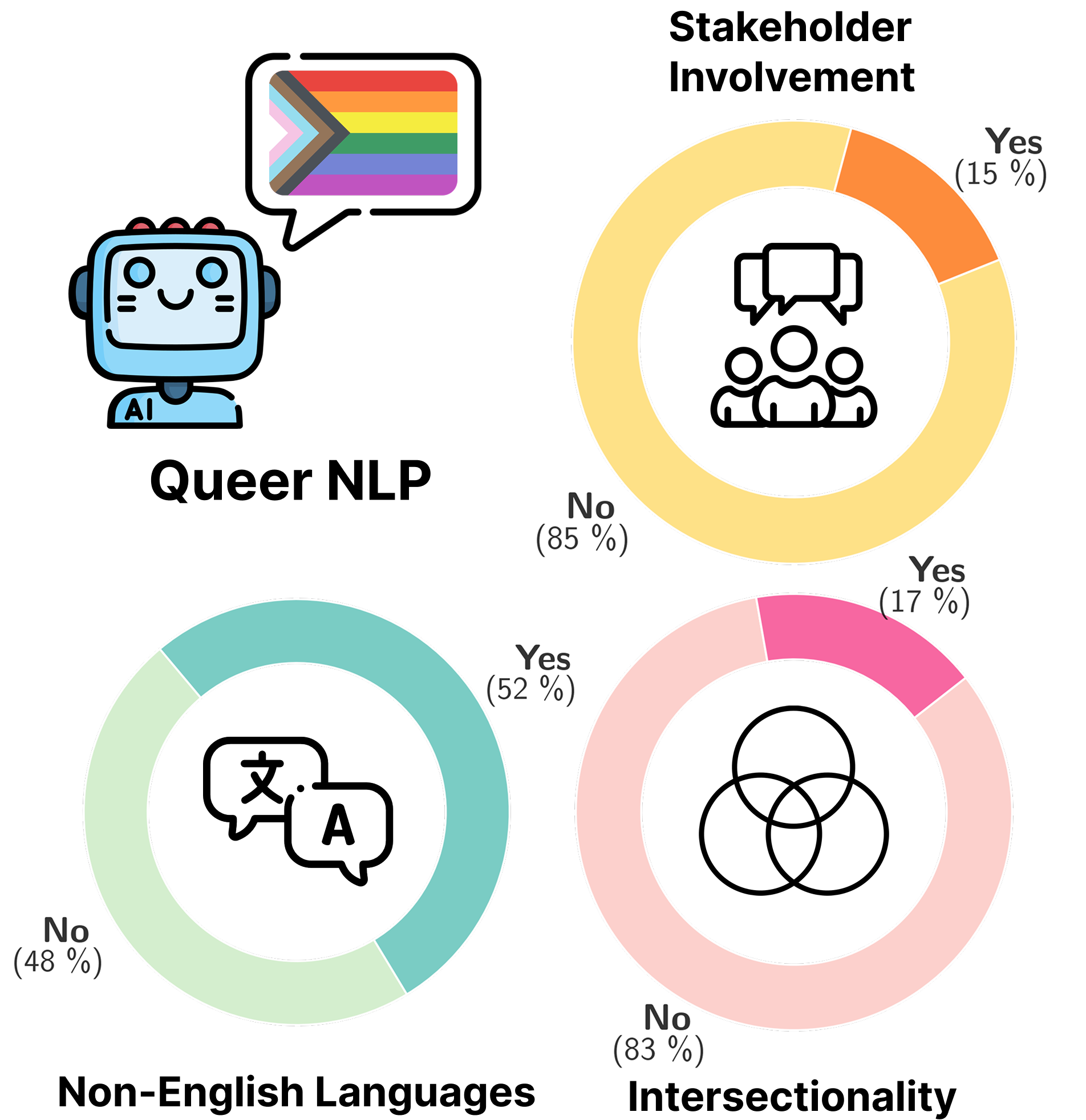}
    \caption{We find that the majority of Queer NLP papers published in the ACL Anthology disregard intersectionality and omit stakeholders. Slightly more than half of all papers include languages other than English. These are descriptive findings, and we provide recommendations in \S\ref{sec:discussion}.}
    \captionsetup{justification=raggedright,singlelinecheck=false}
    \label{fig:findings}
\end{figure}

Our main findings (cf.~\S\ref{sec:common-themes}) are:
\begin{enumerate}
    \item \textbf{Trends:} While there is increased interest in Queer NLP, approaches are often guided by the affordances of current models, focusing on pronouns and hate speech as prominent topics~(\S\ref{sec:common-themes-objects}) and template-based approaches and data augmentation as prominent methods~(\S\ref{sec:common-themes-methods}).
    \item \textbf{Gaps:} Approaches rarely involve stakeholders or take an intersectional view. Nearly half of all papers focus only on English (Figure~\ref{fig:findings}; \S\ref{sec:discussion}). 
    \item \textbf{Future directions:} Beyond closing these methodological gaps, we call on the field to address larger structural gaps, such as dynamic evaluations which allow for temporal and contextual differences, ongoing stewardship of language technologies by the affected communities and a stronger involvement with other disciplines like sociology, gender studies and queer studies~(\S\ref{sec:discussion}).
\end{enumerate}

\vspace{-0.5\baselineskip}
Natural language processing (NLP) is a field concerned with enabling computers to analyze, understand, and generate human language. As NLP technologies  become increasingly integrated into everyday life, users' direct interaction with them has expanded. It is critical that these systems function well, do not reinforce social harms, and maintain the privacy of all users, including queer individuals, who have historically been excluded from technological development and research \citep{techleaversstudy2017, queerinai2023}.

\section{Related Work}
\label{sec:related-work}

The concepts of gender and sexual orientation are central to Queer NLP and have been examined in a number of prior studies.
\citet{devinney2022theories} survey 200 articles on gender bias in NLP to examine how the field conceptualizes gender, \citet{cao-daume-iii-2020-toward} review 150 articles on gender in co-reference resolution. \citet{hobbs2025theoriessexualitynaturallanguage} surveys 55 papers on sexuality bias in NLP, while \citet{Zhou_2024} offers a broad overview of queer harms and structural barriers in NLP. Our work extends these previous approaches with an in-depth survey of the field, spanning all of the named topics and adding ACL anthology papers from different tasks as well, e.g. from machine translation and speech tasks. 

Outside of Queer NLP, several papers survey research on social groups and issues in NLP, including gender bias \citep{stanczak2021surveygenderbiasnatural} and racial bias \citep{field-etal-2021-survey}.
Others survey specific terms, aiming to investigate how they are used in the field, e.g., ``low-resource'' \citep{nigatu-etal-2024-zenos}, ``intersectionality'' \citep{10.1145/3600211.3604705, wang2022intersectionality},  and ``democratization'' \citep{subramonian-etal-2024-understanding}. To our knowledge, our paper is the first to systematically survey Queer NLP papers within the ACL Anthology. 
%

\section{Methodology}\label{sec:methodology}

\subsection{Definition of Queer NLP}
\label{sec:def}
For our purposes, ``Queer NLP'' describes published research work that examines the relationship between the LGBTQIA+ community and NLP technologies. Queer NLP examines how language technologies operationalize concepts related to gender and sexual orientation.

Queer NLP is not simply the application of general bias detection and mitigation techniques to a demographic group. While Queer NLP draws on and contributes to broader fairness and bias methodologies, it addresses modeling challenges that do not easily translate to other groups~\cite{10.1145/3715275.3732033}. It can draw on findings from disciplines like gender studies and queer studies and span multiple levels of linguistic representation. Queer NLP includes phonetic phenomena such as the transcription of queer speakers' voices, syntactic phenomena such as pronoun usage, pragmatic phenomena such as hate speech and reclaimed slurs, and discourse-level phenomena such as code-switching and the co-construction of identity through interaction over time. Consequently, Queer NLP constitutes a distinct and multifaceted area of inquiry focused on phenomena that are linguistically diverse, context-dependent, and resistant to simplification.

\subsection{Paper Selection}

\begin{figure}[t]
\includegraphics[width=\linewidth]{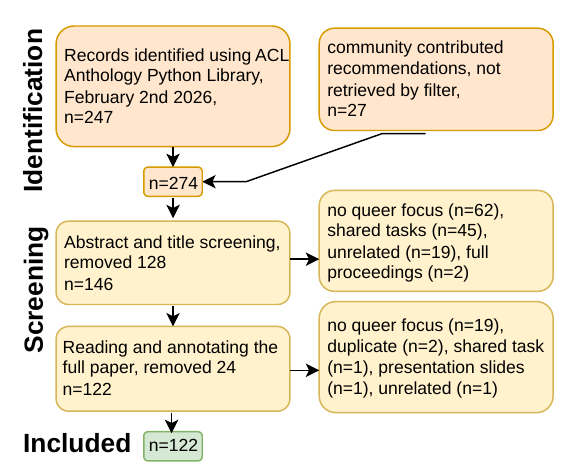}
    \caption{Selection process and filtering for the used papers}
    \label{fig:prisma_figs}
\end{figure}

\paragraph{Selection Criteria.} We intentionally limit our scope to papers 
that place queer people at the center of their inquiry. In this spirit, we only include NLP research that explicitly engages with queerness as a core axis. We exclude work where the engagement with queerness is not a substantial part of the paper, e.g., when queer identity terms are used only as an illustration for a wider phenomenon, queer users are only one of many examined demographic subcategories, or where gender inclusivity is only applied to a binary conceptualization of gender.

We also constrain our paper selection to papers published in the ACL Anthology. While NLP papers are published in other venues as well, the ACL Anthology represents a significant proportion of the discourse in NLP. We do explicitly include papers published at workshops, because research on marginalized communities is, despite its valuable contributions, often not represented at mainstream venues.

\paragraph{Selection Method.} To operationalize these principles we take the following steps (Figure~\ref{fig:prisma_figs}): We use the ACL Anthology Python library\footnote{\href{https://acl-anthology.readthedocs.io/py-v1.1.0/}{\texttt{acl-anthology.readthedocs.io/py-v1.1.0}}} to retrieve papers that feature one or more keywords from a predefined list of queer-specific terms (collected by us) in either their title or abstract. We hypothesize that papers that focus on queer topics would name them prominently. The list contains a subset of the terms used by \citet{10.1145/3613904.3642494}: {\em queer, aromantic, gender non*, lgbt*, agender, glbt, lesbian, gay, bisexual, transgender}, but also words related to systemic oppression such as \textit{heteronormative}, \textit{homophobia}, and \textit{transphobia} and words describing queer-specific harm mitigation such as \textit{gender inclusive} and \textit{gender-fair}. The full list can be seen in Appendix~\ref{appendix:code}. The retrieval was performed on February 2nd 2026, covering all papers in the ACL Anthology previous to this date. This yields 247 papers.

We also allow community contributions provided by us, the authors of this survey, covering papers that could fit our selection criteria but were not retrieved by our search term list. Examples of this category are a paper concerned with neopronouns \cite{gautam-etal-2024-robust}, which are often used by non-binary people, or a paper on a chatbot for education about HIV/AIDS \cite{brixey-etal-2017-shihbot}, which disproportionally affects queer people. This step contributes 27 papers.

We then apply two layers of filtering to check if the papers match the selection criteria. The first filter consists of the manual reading of the titles and abstracts of the papers. This step removes 128 papers, most of them because they lack a queer focus or are unrelated (e.g., when the word `non-binary' describes the property of a graph and not a gender identity). We also remove papers that are submissions to shared tasks, as these would inflate paper numbers while covering the same data set and task. The second filtering is performed during the full-paper annotation step (\S\ref{sec:annotation-process}) and removes 24 papers. Our final survey covers 122 papers.

Questions about whether a paper counts as ``Queer NLP'' are resolved through discussion among us, the authors, using the outlined selection criteria. While we make statistical claims only about the papers contained within the ACL Anthology, our combined expertise brought up many papers that are preprints or published in other venues, which we cite when their perspective is necessary to gain a full understanding of the field. We present all papers as a living resource in an open-source repository, which also contains a file with our annotations.%
\footnote{\url{https://github.com/webersab/queer_NLP_bib}}

\subsection{Annotation Process}
\label{sec:annotation-process}

We annotate each paper in our survey along several axes. First, we annotate the NLP task (e.g., question answering), the motivation, and the method. Second, we annotate which queer groups are explicitly named and what kinds of harm are addressed (if mentioned) and limitations named in the paper. 

We focus on three themes: language diversity, intersectionality and stakeholder involvement. We document the language (see Figure \ref{fig:language_distribution}) as well as the geographical region and genre of the data used. 

We also annotate whether intersectionality is explicitly addressed. Intersectionality is a critical framework combining inquiry and praxis to understand how intersecting systems of power reinforce social inequalities \citep{collins2020intersectionality, crenshaw1989demarginalizing}. In practice, this could include overlapping forms of discrimination, such as harms uniquely affecting Black trans women that differ from those faced by white trans women or Black cis women. We examine how papers account for this complexity and contextualize queer bias and harms within broader systems of power.

Our third focus theme is the explicit involvement of any stakeholders. While work in NLP might not always require stakeholder involvement, there is a growing awareness within the field for the necessity of this approach when dealing with marginalized groups (see \citealp{caselli-etal-2021-guiding} and \citealp{dinkar-etal-2026-can}).

Each paper is read and annotated once, but to calculate inter-rater reliability we randomly select 11 papers (9\% of all collected papers) to be annotated by a second person. We report high reliability across our three focus themes of language diversity, intersectionality and stakeholder involvement, with perfect agreement for language diversity and 90.09\% agreement for the other themes (Cohen's Kappa = .62 for intersectionality and .79 for stakeholder involvement) .

\section{Findings: Common Themes}
\label{sec:common-themes}

Surveying Queer NLP research, we find common trends in what problems are targeted. We also find trends in the methods used. In this section we first look at phenomena that concern queer communities that are reflected in language (\ref{sec:common-themes-objects}) and then examine the methods used both for analysis and mitigation (\ref{sec:common-themes-methods}). While we focus on papers retrieved in our survey, sometimes papers from outside the ACL Anthology can provide useful insights. To distinguish these papers we add a * to their citation in this section.

\subsection{Objects of Study}
\label{sec:common-themes-objects}

\subsubsection{Pronouns}
Pronouns are closely connected to the queer community because they function as a linguistic mechanism through which individuals signal, negotiate, and affirm gender identity in languages where pronouns carry gender information. Gender-neutral pronouns (especially in languages where gendered pronouns are considered the default) and neopronouns feature prominently in our survey. Under-representation of specific pronouns (e.g. English ``xier'') and pronoun–noun combinations (e.g. ``her wife'') in training data has been shown to impair performance in coreference resolution and translation tasks \citep{stewart-mihalcea-2024-whose,friidhriksdottir-2024-genderqueer}. 14 of our papers (11\%) investigate these issues. 

Papers evaluating coreference resolution highlight the harms of misrecognizing people's genders, disconnecting someone's name from their pronouns, or linking their name to a different pronominal chain.
Most papers name trans and non-binary people as the targets of potential harm \citep{dev-etal-2021-harms}, while \citet{cao-daume-iii-2020-toward} state that cis people can also be harmed by misgendering, and \citet{gautam-etal-2024-winopron} argue from a standpoint of universal design (i.e., designing systems to be usable by the widest range of people)  without naming specific communities.

The machine translation (MT) community looks at this problem in regard to translating between languages that mismatch in the gender marking of terms. This work is often found under the term ``gender-fair language.'' and comprises both gender-inclusive and gender-neutral language \citep{paolucci-etal-2023-gender}. At first glance, it is often unclear which papers rely on a binary conception of gender and which explicitly incorporate gender-neutral terms used by trans and non-binary people.
Several papers focus on neopronouns: For example,
\citet{piergentili-etal-2024-enhancing} introduce NEO-GATE, an extension of the GATE benchmark (\citeauthor{10.1145/3600211.3604675}, \citeyear{10.1145/3600211.3604675}*), for translation from English to Italian, and \citet{piergentili-etal-2025-llm} adapt chain-of-thought prompting to improve gender-neutral translation. \citet{lauscher-etal-2023-em} incorporate nounself, emojiself, numberself, and other neopronouns in addition to gender-neutral and gendered pronouns while evaluating MT models.

Another way to create gender-fair texts is rewriting and post-editing:
\citet{lardelli-gromann-2023-gender} recruit translators to post-edit English to German machine translations into gender-fair language, while \citet{veloso-etal-2023-rewriting} create a rule-based and a neural-based rewriter for Portuguese. \citet{amrhein-etal-2023-exploiting} use round-trip translation to create pairings of unbiased texts with biased translations to train a de-biasing rewriting system. 

\paragraph{Recommendations} Resolution of neopronouns or non-normative combinations of pronouns and nouns is one exemplary case in which human performance easily surpasses state-of-the-art machine learning models, serving as a litmus test for when a machine learning model is memorizing surface forms rather than learning underlying rules. This makes it a phenomenon that is easily measurable and therefore shows up often in NLP publications. Nevertheless, how downstream performance affects queer users remains under-examined. Other examples of research gaps in this area are the treatment of context-based switching between pronoun sets and the usage of several different pronoun sets for the same person. A way forward is increased stakeholder involvement, as described by \citet{gromann-etal-2023-participatory} who study how trans and non-binary people are impacted by machine translation and what their wishes for better systems are.

\subsubsection{Hate Speech and Slurs} 

In recent years, detecting transphobic and homophobic hate speech has become a prominent focus, with numerous shared tasks addressing this challenge across multiple languages 
(\citeauthor{chakravarthi-etal-2022-overview}, \citeyear{chakravarthi-etal-2022-overview}, \citeyear{chakravarthi-etal-2023-overview}, \citeyear{chakravarthi-etal-2024-overview}; \citeauthor{overview-hodi2023}, \citeyear{overview-hodi2023}; \citeauthor{overview-homomex2023}, \citeyear{overview-homomex2023}*; \citeauthor{overview-homomex2024}, \citeyear{overview-homomex2024}*; \citeauthor{overview-hope2023}, \citeyear{overview-hope2023}*; \citeauthor{overview-hope2024}, \citeyear{overview-hope2024}*). 34 of the papers in our survey (28\%) examine hate speech.

When trained on data stripped of its situational context, systems frequently misclassify identity terms as slurs or negative expressions.
This leads to problems in hate speech classification, where systems over-classify instances with queer terms as offensive, while not flagging offensive text that lack these words.
Several papers show that models regularly misclassify words like {\em gay, trans,} or {\em queer} as toxic, even when used neutrally or positively \citep{zueva-etal-2020-reducing, sahoo-etal-2022-detecting}.
Some papers show that text classification systems are also biased against language used among queer people \cite{wu-hsieh-2017-exploring, ramesh-etal-2022-revisiting} and references to queer identities \cite{tint-2025-guardrails, ungless-etal-2023-stereotypes, zhang-etal-2020-demographics, zueva-etal-2020-reducing}.

Despite the prevalence of hate speech in a diversity of languages, a relatively small number of papers study languages other than English.
Work has been done on Mexican Spanish \cite{vasquez-etal-2023-homo}, Russian \cite{zueva-etal-2020-reducing}, Malayalam and Tamil (\citeauthor{10.1007/978-3-031-28183-9_15}, \citeyear{10.1007/978-3-031-28183-9_15}*), and Gujarati, Kannada, and Telugu \cite{kumaresan-etal-2024-dataset}. \citet{locatelli-etal-2023-cross} study homotransphobic speech on Twitter in seven languages, concluding that current automated moderation systems are insufficient and have large differences across languages.

\paragraph{Recommendations} Online content moderation is one of the spaces where queer users can directly feel the impact of language technologies by being silenced, censored or targeted, which makes it an impactful field of study. In the case of hate speech, who is speaking to whom is just as important as what is being said, but there is a gap in research on in-group and out-group language (one notable exception being \citet{draetta-etal-2024-reclaim}). However, detecting in-group language can be a double-edged sword, as machine learning systems that recognize in-group usage of queer language to prevent censorship can be used to expose queer people who use coded language to escape detection. We recommend the community focus on the tension between detecting in-group language use and safeguarding those who rely on it for privacy and safety. Also, current methods have focused primarily on flagging harmful speech and less on promoting counterspeech. Notable exceptions are works on hope speech which have started to appear in the NLP ethics community, including in Queer NLP \citep{pofcher-etal-2025-hope}. 

\subsubsection{Stereotypes}
Besides the category of hate speech, queer people are represented in the public discourse in many different ways. Stereotypes of queer people are propagated across social and traditional media, their presence or absence reflecting localized climates of acceptance or hostility. When stigmatizing stereotypes become training data, they can be reproduced by NLP systems, resulting in harmful biases that surface in downstream applications. In this sense, stereotypes can be studied both in the underlying training data and in the analysis of LLM outputs, which opens its own field of study that is distinct from the examination of hate speech and slurs. 

Some works use information extraction tools to evaluate texts written about the queer community. \citet{locatelli-etal-2023-cross} and  \citet{andersen-etal-2024-mexican} analyze English and Mexican Spanish homotransphobic speech on Twitter, respectively, noting a decrease in the use of many derogatory words over time, while the overall vocabulary for queer terms has increased.
\citet{ch-wang-jurgens-2021-using} also study Twitter and Reddit data, finding that the use of gender-neutral expressions (e.g., partner, spouse, folks) has risen steadily, with the sharpest uptake among heterosexual, liberal, and higher-socioeconomic-status users.
\citet{hicks-etal-2016-analysis} use Twitter data to find terms for gender expression that do not appear in the US National Transgender Discrimination Survey to improve survey design.
These papers demonstrate the potential of NLP tools in analyzing the use of queer terms in society more broadly, leading to increased knowledge of the popular discourse about queer people.

When it comes to queer representation in the outputs of LLMs, there has been a wide application of template-based methods to elicit stereotypes, which we cover in more detail in \S\ref{sec:common-themes-methods}. 

An important aspect of stereotypes is their changeability. \citet{Zhou_2024} draws attention to the fact that understandings of queerness vary both across cultures and over time.
Given the fluid nature of ``queerness'' as used by queer people and noted by philosophers \citep{butler2020critically}*, models of anti-queer bias must be equally fluid to account for different dataset contexts.
For instance, inoffensive or clinical terms from the past, such as ``transvestite'', can gain negative associations through pejoration \cite{finkbeiner2016pejoration}*.

Representation of queerness in NLP technology would hence require constant rewriting by the affected community, similarly to the slow and constant change of language in different communities of practice, which the affordances of current NLP technologies make hard to imagine.

\paragraph{Recommendation} It remains important to examine how queer people are discussed online and how ongoing language change shapes the terms associated with them. LLM outputs can be viewed as a reflection of such discourse, while also revealing these systems’ tendency to internalize latent social patterns, including hetero- and cisnormativity. \citet{zhou-etal-2025-culture}* point out that LLMs are only capable of reproducing discourse about social groups, rather than capturing how social identities are enacted through language. Looking into this topic is in line with the study of in-group language and would shine a light not only the most overt manifestations of bias but also on subtle and context dependent patterns of normativity and erasure. 

\subsubsection{Underrepresentation}
In many curated datasets, queer data are underrepresented or even filtered in fear of being offensive \citep{dodge-etal-2021-documenting}. Lack of queer-specific data can then lead to diminished service performance and harms to queer users. 

Speech processing is one domain in which queer representation remains limited.\footnote{ACL affiliated conferences have been explicitly welcoming speech papers, yet none of the ACL Anthology venues appear to be the places where people choose to publish their works on queer speech tech. Subsequently, most of the papers covered in this section are not included in the 122 papers retrieved from the ACL Anthology, but can be found in the accompanying GitHub repository.} 
Processing spoken language is relevant for transgender and non-binary voices, since they challenge normative assumptions about gender (\citeauthor{zimman2028transgender}, \citeyear{zimman2028transgender}*, \citeyear{zimman2021gender}*). Nevertheless, speech technology research focuses on binary genders and cisgender speech (\citeauthor{sanchez2024binary}, \citeyear{sanchez2024binary}*).
Voice also affects the expression and perception of sexual orientation 
(\citeauthor{levon2007sexuality}, \citeyear{levon2007sexuality}*; \citeauthor{simpson2020phonetic}, \citeyear{simpson2020phonetic}*).

Very few speech datasets explicitly contain speech by queer individuals, although such data might be included and simply not labeled.
Although this hinders research on the robustness of speech tools, publicly sharing such metadata comes with privacy risks.
\citet{siegert25queerwaves}* present a German speech dataset of podcasts and videos by people who publicly identify as queer.%
\footnote{Three more corpora without corresponding publications focus on transgender and/or non-binary English voices 
(\citeauthor{german2022gender}, \citeyear{german2022gender}*; \citeauthor{dolquist2023palette}, \citeyear{dolquist2023palette}*; \citeauthor{hope-mages}, \citeyear{hope-mages}*).
Another largely non-queer corpus contains metadata on gender expression and sexual orientation (\citeauthor{weirich24gender}, \citeyear{weirich24gender}*).}
An audit of seven major audio corpora finds that queer identities are rarely discussed, and several datasets additionally contain queerphobic content (\citeauthor{agnew2024soundcheck}, \citeyear{agnew2024soundcheck}*).

Only a single project focuses on queer voices in the context of Automatic Speech Recognition: a recurring shared task on ASR for underrepresented social groups in India including transgender people \cite{b-etal-2022-findings-shared}. 
However, none of the overview or participant papers address gender, transness or queerness.

Three papers build speech synthesis systems for voices perceived as non-binary
(\citeauthor{danielescu2023creating}, \citeyear{danielescu2023creating}*; \citeauthor{szekely24_interspeech}, \citeyear{szekely24_interspeech}*; \citeauthor{hope25_interspeech}, \citeyear{hope25_interspeech}*)
using speech data from non-binary and trans speakers, and asking non-binary people to evaluate the resulting systems.
\citet{sigurgeirsson24camp}* examine the shortcomings of voice cloning text-to-speech systems in replicating gay voices.
They also discuss dual-use issues with queer speech technology, including their use for mockery.


Most publications on queer speech technology are very recent (2022--2025).
The special session on Queer and trans speech science and technology\footnote{\href{https://sites.google.com/view/is-queer-trans/home}{\texttt{sites.google.com/view/is-queer-trans/home}}} at Interspeech 2025 also demonstrates a current interest in analyzing and processing queer speech.%
\footnote{Outside of NLP, \citet{netzorg24speech}* focused on how speaker identification systems struggle with voice modulation for gender expression and \citet{doukhan23voice}* and \citet{mcallister25_interspeech}* on systems that help with voice transition.}
Given the (non-queer) research on ASR quality differences based on speaker gender (\citeauthor{feng2024towards}, \citeyear{feng2024towards}*), one future research direction might be to investigate performance gaps for gender-non-normative voices, including in other tasks with speech input, such as spoken language understanding and speech translation.
In speech translation or summarization, queer speakers might be especially prone to being misgendered, which has been acknowledged \cite{gaido-etal-2020-breeding}, but not explicitly investigated.

Another domain in NLP where under-representation plays a pivotal role is research into the accessibility of information on queer topics via chatbots.
\citet{Najafali2023}* investigate ChatGPT's ability to provide evidence-based recommendations for gender-affirming surgery according to the standards of care by the World Professional Association for Transgender Health.
Although ChatGPT can accurately describe aspects of gender diversity and treatment for gender dysphoria, the study highlighted gaps in completeness, reliability, and explicit reference to authoritative guidelines.
\citet{ais.2024.91}* present a similar user study, showing low trust in ChatGPT's ability to provide gender-affirming surgery information. 
Motivated by the high engagement of queer youth in social media, \citet{brixey-etal-2017-shihbot} develop a chatbot for HIV prevention and care for queer youth.
They create a corpus of 3,000 question-and-answer pairs from reliable sources and train a chatbot which they deploy on Facebook Messenger.
However, their work does not analyze their success in reaching queer youth. 

Some works study chatbot use by queer individuals for mental health and emotional support.
\citet{lissak-etal-2024-colorful}, \citet{10.1145/3469595.3469597}, and \citet{10.1145/3613904.3642482} conclude that these systems have positive effects, but are generic and culturally insensitive outside the US.
While unsuitable to replace human emotional intelligence, the authors argue that a chatbot may be a useful tool for initial support.

\paragraph{Recommendations} Our survey shows that speech technology and information access to queer topics are two domains where a lack of queer representation leads to potential harms for queer people. There is a tricky balance when it comes to inclusion and representation in these systems. In some cases inclusion can mean safety, e.g., when accurate queer health information is accessible via broad purpose chatbots. In other cases it could mean endangerment trough outing or the facilitation of surveillance, e.g. when speech recognition systems trained on queer voices disclose that a user is trans. Representation is not a panacea. We therefore recommend thinking of the consequences of inclusion of data or the collecting of gender and sexual orientation information, ideally in close collaboration with stakeholders.

\subsection{Methods}
\label{sec:common-themes-methods}

\subsubsection{Template-based Methods for Bias Detection}
Large Language Models (LLMs), which increasingly serve as the backbone of modern NLP systems, are trained on massive, mostly uncurated online text corpora.
LLMs thus tend to inherit and reproduce social biases from these corpora \cite{mickel2025more}*. Examining LLMs and their outputs is therefor a common object of study in Queer NLP papers (30 papers, 24\% of our survey).

Template-based methods are very prevalent in the examined papers, both for the measuring of bias and for the creation of training data. This can be seen as a product of the present time, as the affordances of LLMs make these methods expedient.
As the most widely-used approach to elicit LLM bias, template-based methods involve prompting LLMs with sentences containing subject placeholders, which can be filled with terms representing various social groups, e.g., identity terms relating to gender and sexuality, names, or pronouns 
(\citeauthor{nozza-etal-2022-measuring}, \citeyear{nozza-etal-2022-measuring}; \citeauthor{dhingra2023queerpeoplepeoplefirst}, \citeyear{dhingra2023queerpeoplepeoplefirst}; \citeauthor{hossain-etal-2023-misgendered}, \citeyear{hossain-etal-2023-misgendered}; \citeauthor{sosto2024queer}, \citeyear{sosto2024queer}*; \citeauthor{ovalle2023}, \citeyear{ovalle2023}*). Notably, \citet{felkner-etal-2023-winoqueer} take a community-centered template-based approach by collaborating with queer annotators to generate sentence pairs, such as ``LGBTQ people are sinful'' versus ``Heterosexual people are sinful''.

By analyzing the outputs conditioned on these different terms, the authors report systematic differences that often reflect social biases or imbalances.
Given the complexity of assessing LLM bias in queer-related contexts, many studies rely on intrinsic bias measures. Building on template-based methods, common approaches include direct statistical comparison between expected and actual answers \cite{bunzeck-zarriess-2024-slayqa, xie-etal-2024-addressing}, as well as  comparison of probabilistic scores such as pseudo-log-likelihood or perplexity for target-attribute pairings within sentence templates
(\citeauthor{felkner-etal-2023-winoqueer}, \citeyear{felkner-etal-2023-winoqueer}; \citeauthor{hossain-etal-2023-misgendered}, \citeyear{hossain-etal-2023-misgendered}; \citeauthor{bergstrand-gamback-2024-detecting}, \citeyear{bergstrand-gamback-2024-detecting}; \citeauthor{xie-etal-2024-addressing}, \citeyear{xie-etal-2024-addressing}; \citeauthor{ovalle2023}, \citeyear{ovalle2023}*).

In contrast, other studies perform group-level comparisons using downstream tasks, yielding more extrinsic bias measures. For instance, sentiment analysis has been widely applied to examine polarity (negative, neutral, or positive) in LLMs outputs
(\citeauthor{kiritchenko-mohammad-2018-examining}, \citeyear{kiritchenko-mohammad-2018-examining}; \citeauthor{sosto2024queer}, \citeyear{sosto2024queer}*),
while regard analysis \citep{sheng-etal-2019-woman} has been used to capture perceived respectfulness in generated text \cite{ovalle2024rootshapesfruitpersistence, dhingra2023queerpeoplepeoplefirst}.
Similarly, harmful language detection tools such as Perspective API\footnote{\url{https://www.perspectiveapi.com}.} are frequently used to detect toxicity and offensive language
(\citeauthor{nozza-etal-2022-measuring}, \citeyear{nozza-etal-2022-measuring}; \citeauthor{sosto2024queer}, \citeyear{sosto2024queer}*), alongside lexicon-based approaches such as HONEST (\citeauthor{nozza-etal-2021-honest}, \citeyear{nozza-etal-2021-honest}*), which builds on HurtLex (\citeauthor{bassignana2018hurtlex}, \citeyear{bassignana2018hurtlex}*) to quantify offensiveness.

Several papers use other tasks as a testbed to assess the harmfulness of LLMs. For instance, \citet{sobhani-etal-2023-measuring} evaluate gender bias in the context of hate speech and offensive language classification. 
In the same vein, some works also make the use of question-answering datasets for this purpose. SlayQA \citep{bunzeck-zarriess-2024-slayqa} provides contexts and question that specifically probe the model's ability to handle neopronouns, while BBQ \citep{parrish-etal-2022-bbq} provides contexts that are intentionally kept ambiguous as to expose LLMs' internal biases about different social groups. This is also looked into by \citet{shaier-etal-2023-emerging}, who analyze how the additional of superfluous demographic information changes the answer of LLM-based medical question-answering systems.

These methods are rarely tailored to capture the specific nuances of queer or intersectional identities in an usage context, which limits their ability to provide a full picture of harmful biases in LLMs.

Another under-examined problem is ``unmarkedness'' in language, as unnatural mentions of privileged identities in templates e.g., ``straight man'' may yield unreliable results \citep{blodgett-etal-2021-stereotyping}.
Here, a lack of engagement with linguistics may lead to insufficient methods.

\paragraph{Recommendations} Introducing yet another dataset or template-based bias elicitation method offers limited additional insight. Contemporary LLMs can readily internalize such benchmarks to strengthen guardrails, which often makes the benchmark obsolete and results in sanitized yet still stereotypical outputs. What is needed instead are novel measurement paradigms that are grounded in real-world usage contexts and a stronger theoretical foundation for operationalizing phenomena such as erasure and heteronormativity, particularly once more overt manifestations of bias have been obscured.  

\subsubsection{Post-hoc Bias Mitigation}
A central motivation in Queer NLP is to mitigate anti-queer biases in models.
However, this is usually restricted to first revealing the existence of some bias, and only subsequently developing or applying methods that mitigate it.
This reactive approach often results in a cat-and-mouse game, where papers demonstrate bias in existing models but are unable to develop better models, as debiasing methods are rarely adopted or feasible to operationalize at scale.
This pattern has also been observed in other subfields of NLP, e.g., in interpretability research \citep{gururaja-etal-2023-build,mosbach-etal-2024-insights,orgad2026interpretabilityactionable}.

The deficiencies of models can often be traced back to their training data, and better resources are thus one potential solution.
For example, \citet{ljubesic-etal-2020-lilah} introduce the LiLaH emotion lexicon for Croatian, Dutch, and Slovene, showing more consistent results on LGBTQ topics in emotion classification.
\citet{waldis-etal-2024-lou} create a resource for German text classification for seven different tasks, including gender-neutral and gender-fair formulations.

In hate speech and toxicity detection, \citet{engelmann-etal-2024-dataset} and \citet{sahoo-etal-2022-detecting} construct datasets to detect dehumanizing language.
\citet{dacon2022detecting} focus entirely on LGBTQIA+ individuals, creating a real-world dataset of harmful online conversational content.
\citet{wiegand-ruppenhofer-2024-oddballs} provide a dataset and train predictors for more subtle language where a group is portrayed as deviating from a norm.
In a similar vein, \citet{lu-jurgens-2022-subtle} introduce a dataset to detect trans-exclusionary language by TERFs (see \hyperref[sec:glossary]{Glossary}).
\citet{10.1145/3658644.3670284} create a debiasing dataset using Counterfactual Data Augmentation (CDA) to de-bias biased contexts.
They construct anti-biased descriptors, i.e., replacing stereotypical associations with counter-stereotypical ones to generate texts with them, creating a dataset for fine-tuning.

On the side of de-biasing interventions that go beyond data, \citet{dhingra2023queerpeoplepeoplefirst}* apply a post-hoc method that combines chain-of-thought prompting with SHAP analysis (\citeauthor{10.5555/3295222.3295230}, \citeyear{10.5555/3295222.3295230}*) to identify words in the LLM output that reduce social regard. After detecting low-regard words, they query an LLM for explanations and then re-prompt it to rephrase the sentence, removing those words while preserving meaning. 

\paragraph{Recommendations} Bias mitigation is an important part towards safe and fair NLP systems. However, rather than focusing solely on existing systems and their shortcomings, we recommend researchers to envision and co-create language technologies that are not merely harmless but actively affirming of queer identities and supportive to queer users. Grassroots efforts outside of academia exist in this regard, e.g., Shinigami Eyes\footnote{\url{https://shinigami-eyes.github.io/}}, a browser extension that highlights transphobic and trans-friendly social network pages in different colors to help queer users navigate to safe online spaces. Bringing stakeholder ideas and use cases into research practice can prove as a fruitful way forward. 

\section{Current Gaps and Future Work}
\label{sec:discussion}
By characterizing common phenomena and methods, we also outline the gaps of existing research, some of which are mentioned in the cited papers as avenues for future work. In this section, we examine them and point to positive exemplars of first steps towards closing them, before moving on to more overarching topics. 

\subsection{Cis-normative conceptions of gender}
An important thread of discussion in Queer NLP papers concerns the treatment of gender in research. 
\citet{devinney2022theories} criticize the field's predominantly binary and cis-normative conceptions of gender.
\citet{hovy-spruit-2016-social} note that researchers often use pronouns or names as a proxy for gender, and \citet{gautam-etal-2024-stop} extend this observation to other sociodemographic attributes, describing ethical issues and giving practical advice.
The framing of the treatment of gender data as an ethical issue has already been argued before by \citet{larson-2017-gender}, outlining ethical principles about the use of gender, and
suggesting that researchers make underlying theory of gender explicit and use gender only if necessary. 

We would like to see these recommendations more widely adopted in NLP, especially when papers address gender bias. Gender bias dedicated venues such as the Workshop on Gender Bias in NLP\footnote{\url{https://gebnlp-workshop.github.io/}} should release a set of guidelines for the treatment of gender as a variable (a ``gender bias starter pack'') to both guarantee a high standard within its accepted publications and to facilitate the permeation of best practices throughout the community.

\subsection{Stakeholder Involvement}
Although most papers name impact on the queer community as a motivating factor for their research, very few include queer community members as part of the development or evaluation of their methods\footnote{While not a substitute for broad stakeholder engagement, Queer NLP research is often conducted by queer researchers drawing on lived experience. These researchers face a double bind: disclosing their positionality (see Appendix \ref{sec:ethics}) can strengthen the grounding of their work, but may also risk negative career consequences through being outed.}.
More often, linguistic representations like identity terms, pronouns, and automatically-calculated metrics are used as a stand-in for users.
Notable exceptions are: \citet{gromann-etal-2023-participatory}, who conduct a workshop with representatives from the intersex and non-binary communities to determine how they would like to be represented in translation; \citet{felkner-etal-2023-winoqueer}, who create the WinoQueer Benchmark with text input from the community; and \citet{ungless-etal-2023-stereotypes}, who survey trans and non-binary people on their experience with generated images, specifically asking for ways in which users would like to be represented. Nevertheless, these approaches engage with the stakeholders only once and do not establish ways in which the community is continuously involved in the stewardship of technologies.

There are two promising avenues through which Queer NLP can improve in the future. First, recent work on low-resource and endangered languages has increasingly emphasized stakeholder involvement (e.g., \citealp{bird-yibarbuk-2024-centering}), offering a useful model for Queer NLP. Second,  relevant work on queer stakeholder involvement is likely also taking place in human--computer interaction venues, rather than in papers from the ACL Anthology, which primarily captures language technology interventions. Thus, future NLP research could benefit from drawing on findings from human--computer interaction literature.

\subsection{Intersectionality and Interdisciplinarity} \citet{10.1145/3600211.3604705} point out that AI fairness literature mainly engages with intersectionality when applying metrics to demographic subgroups, and avoids other aspects, such as examining power structures.
Similarly, \citet{devinney-2025-power} argues that in order to mitigate the negative effects of stereotyping, NLP researchers need to engage with the social mechanisms behind it. However, we find that only 17\% of all examined papers explicitly mention intersectionality.

\begin{figure}[t]
    \centering
    \includegraphics[width=\linewidth, clip]{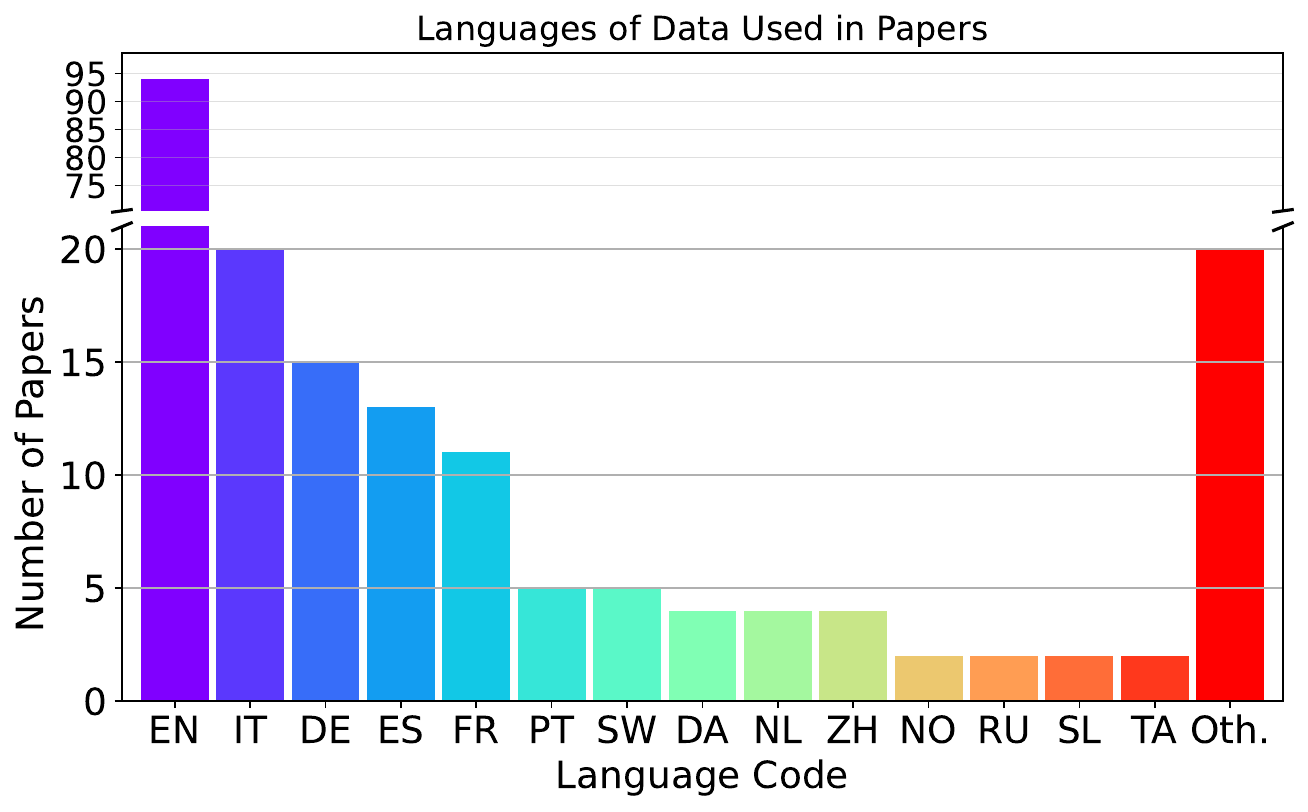}
    \caption{Breakdown of the 35 languages represented across papers. The `Other' (Oth.) category aggregates all languages that appeared in only one paper. See Appendix \ref{languages} for a full list of languages along with paper counts.}
    \label{fig:language_distribution}
\end{figure}

We also find that the selection of studied languages limits the opportunity to engage with a variety of intersecting identities in the global queer community. Slightly less than half of the surveyed papers examine exclusively English language data, and even the inherently multilingual field of machine translation considers only translations to and from English (we examine non-English Queer NLP more closely in \S\ref{sec:non-english-queer-nlp}). Languages other than English skew towards Romance and Germanic languages (see Figure \ref{fig:language_distribution}), which leads to a lack of cultural diversity and a concentration on concepts of queerness prevalent in the Global North. 

Operationalizing intersectionality requires new data collection and processing practices and a deep engagement with the communities that they touch on, going beyond the currently practiced one-off involvement of stakeholders towards an integration of them in every step of the research process. 

NLP can be thought of as an inherently interdisciplinary field that straddles linguistics and computer science.
Especially when dealing with queer language phenomena, previous work in sociolinguistics, gender studies, and queer studies can provide guiding principles for research.
One positive example is \citet{amironesei-diaz-2023-relationality}, who suggest modeling social relations as a pathway to improved hate speech detection.

While technological affordances shape much of what we see in our survey, we recommend researchers engage with the rich existing literature in the aforementioned fields to arrive at new ways of approaching problems and at better ways to connect social phenomena with data and model artifacts. 

\subsection{Non-English Queer NLP}\label{sec:non-english-queer-nlp}

An inherent limitation of our survey is the focus on ACL.
Even with its international appeal, the kind of paper submitted and accepted is still geared towards an English-speaking audience \citep{lepp2024global}, which leads to an unequal distribution of target languages (see Figure \ref{fig:language_distribution}).

The predominance (if not hegemony) of English decreases when we broaden our scope beyond the Anglosphere.
The Brazilian Computer Society has hosted workshops featuring papers on Queer NLP about and written in Portuguese \citep{bresci, brasnam}, and so has the Italian Association of Computational Linguistics, although written in English \citep{overview-hodi2023, draetta-etal-2024-reclaim}.
The Spanish Society for Natural Language Processing is particularly prolific, with many Queer NLP papers on Spanish and other Iberian languages at their IberLEF workshop \citep{overview-homomex2023, overview-hope2023, overview-homomex2024, overview-hope2024}.

These workshops address topics that could be explored in English Queer NLP papers but are currently underrepresented.
For example, \citet{Garcia-Baena2023-ic} introduce the task of hope speech detection for queer individuals in Spanish, defined as the detection of ``speech that is able to relax a hostile environment and that helps, gives suggestions and inspires for good'', which culminated in HOPE, a shared task on hope detection at IberLEF.
The first edition in 2023 focused on queer hope speech \citep{overview-hope2023} and the second edition in 2024 still dedicated half of the task to it \citep{overview-hope2024}. While this topic is less explored in English Queer NLP, a notable exception is \citet{pofcher-etal-2025-hope}.
Similarly, the study of the identification of reclaimed queer slurs in Italian \citep{draetta-etal-2024-reclaim} spawned a shared task on multilingual queer slur reclamation for the EVALITA workshop 2026, \footnote{\url{https://multipride-evalita.github.io/}} a topic that remains understudied in English NLP \citep{10.1145/3614419.3644025}.

Shared tasks play a key role in these events, drawing researchers' attention to the subject.
The HOPE tasks from above included 8 papers in 2023 \citep{overview-hope2023}
and 16 in 2024 \citep{overview-hope2024},
making it the largest repository of papers on hope detection for the queer community.
There are similar numbers for hate detection in Mexican Spanish with the HOMO-MEX task at IberLEF \citep{overview-homomex2023,overview-homomex2024} and in Italian with the HODI task at EVALITA \citep{overview-hodi2023}.
This may indicate that specific challenges help gather momentum for research in non-English Queer NLP, which can be especially useful for identities that don't fit the Western mold of ``queer'' e.g., hijra (see \hyperref[sec:glossary]{Glossary}).

Much like English Queer NLP research influences the papers at venues not focused on English, many of their ideas and advances have relevance for languages beyond the one under study. We encourage researchers to broaden their focus accordingly.

\subsection{Harms in the Wild vs. in Research}\label{sec:harms-in-the-wild-and-research}

Most Queer NLP research has focused on a relatively narrow set of harms.  Reliance on rigid taxonomies (e.g., of harm types, speech types, and identities) can flatten the queer experience 
 while disregarding the fluidity of queer language and the necessity for expert stakeholder input.
Creating these taxonomies is a fundamentally political decision that influences what is visible and invisible within a system \cite{bowker2000sorting, smith2018becoming, GuyanRainbowTrap}. Most hate speech and toxicity detection systems, for example, categorize (implicitly or explicitly) their targeted speech, but 
rarely involve the perspectives of queer users who experience hate speech: only 4 out of the 34 papers surveyed involve queer stakeholders. 

Classification comes with potentially dire real-world consequences for queer people, given that queerness---as a deviation from what is constructed as ``normal'' or desirable---has historically been an important target of surveillance \citep{Kafer_Grinberg_2019}, ostensibly on grounds of public health, reproduction, and morality \citep{Conrad_2009}.
Surveillance has an outsized effect on the most marginalized in our societies \citep{Eubanks2018-EUBAIH}, and is further enhanced by several pursuits in NLP, including entity tracking, event tracking, sentiment analysis, and social media monitoring, all of which we have touched on in this survey.
However, despite the way that NLP systems enable surveillance, there is little work that calls out these risks, let alone critiques our role as researchers in creating them \citep{kaffee-etal-2023-thorny}.

In order to avoid the negative consequences of both (mis)classification and surveillance, queer users may reject the premise that systems should categorize them at all.
Critical refusal, or opting out of being measured, labeled, or represented in normative ways, is a vital queer methodological position that current NLP systems leave little room for \citep{abramovich-ma-2024-evaluating-refusal}.
Refusal is not simply a rejection of inclusion, but a rejection of the frameworks that inclusion demands: that identities be stable, categorizable, and intelligible to institutions \cite{costanza2020design, crawford2021atlas}.

In practice, this could include actions such as declining to specify a gender on a form, using intentionally ambiguous language online, resisting classification through irony, camp, or coded speech, or, among NLP researchers, refusing to use harmful categorization in research~\citep{gautam-etal-2024-stop}. 
These modes of resistance challenge NLP systems that depend on fixed typologies and clear labels, exposing a gap between what queer people want from NLP technologies and what those technologies are designed to do.
However, refusal is difficult to publish about in NLP, as it definitionally resists detection, annotation, and evaluation.
As a result, queer refusal is rendered technically invisible even as it remains politically relevant.

\subsection{Perspectives from Queer Theory}\label{sec:queer-gender-studies}

One aim of this survey is to look beyond immediate gaps in Queer NLP and to create visions for future work.
To understand the complexity of socially-constructed identities and their interactions with sociotechnical systems, we can turn to queer theory approaches. This work examines and challenges power \citep{foucault1978history} while confronting the certainty of sex, gender, and sexuality \citep{butler_gender_1990}.
Acknowledging that power plays a role in shaping what it means to be ``queer'' can help us better conceptualize why queer life is often denigrated or outright erased in NLP systems, providing a path forward for better Queer NLP research.

One of queer theory's aims is to question the stability of identity categories, taking particular issue with how these categories are presumed and normalized in everyday life and academic discourse.
NLP is currently no less guilty of adhering to strict categorization, particularly when it comes to encoding queerness.
In our survey, all papers that specifically study queerness rely on lists of sexual and gender identities, reducing nuanced or fuzzy forms of queer life into simple categorizations.
Further, \citet{klipphahn-karge_introduction_2023} argue that AI ``posits societal norms within a double bind'' by both \textit{reflecting} existing social bias and \textit{producing} heteronormative knowledge that flattens multiplicities. 
Because of this, some queer theorists claim that data science and its applications, such as NLP, are entirely incompatible with queerness \citep{keyes2019counting}.
Others, such as \citet{shah_i_2023}, propose ways to ``queer'' AI systems by reworking data science building blocks to allow imperfect, non-normative, and community-gathered data points.
In either case, in order to break down cis- and heteronormative NLP standards, we must expand our scholarship and collaborations from the intersection of computing, statistics, and linguistics to include sociology, gender studies, queer theory, and beyond.
With one foot ``in the craft work of design'' and the other ``in the reflexive work of critique'' \citep{agre_toward_1998}, we can build interdisciplinary systems that proactively center those at the margins instead of relying on post-hoc tweaks and retroactive guardrails.

\section{Conclusion}\label{sec:conclusion}

This work provides an overview of the current state of Queer NLP, summarizing findings from a variety of subtopics, as well as contextualizing them~(\S\ref{sec:common-themes}--\ref{sec:discussion}). Limitations and our ethics statement can be seen in Appendix \ref{sec:limitations} and \ref{sec:ethics}.

Queer NLP echoes many shortcomings of the field as a whole.
Despite some recent efforts (\S\ref{sec:non-english-queer-nlp}), slightly less than half work focuses exclusively on English, with most of the work centering English even when other languages are included (e.g., in machine translation).

Moreover, many works operate with a binary and cis-normative viewpoint of gender and sexuality, which could be informed better by research from queer theory.
Finally, any harm mitigation usually happens post-hoc;
this can at least partially be explained by the lack of stakeholder involvement in modern NLP systems.

For these reasons, we advocate for future work to radically rethink the ways in which we build language technologies. Going forward, researchers should develop solutions for socially-grounded rather than assumed understandings of harm and therefore involve affected queer communities. 
Community involvement needs to go beyond one-off workshops and data collection efforts towards an integration into all steps of the research pipeline, from the definition of use cases over model development to long-term ownership and auditing of resulting systems.

Our way forward can take inspiration from research on low-resource languages and human-computer interaction, through community-driven data collection and participatory action research, which involves stakeholders as collaborators and challenges oppressive epistemologies and methods. Moreover, interdisciplinary collaboration with researchers of sociology, gender studies, and queer theory can open new avenues to move beyond restrictive methods of data collection and model building.
We call for the development of NLP systems that not only mitigate harm but actively support and uplift queer communities to make them the stewards of the language technologies that affect them.


\section*{Acknowledgments}
We would like the reviewers and the action editor for their helpful comments. We would also like to thank all members of Queer in AI who supported us with advice throughout the whole process of writing this paper.
\bibliography{custom}
\bibliographystyle{acl_natbib}


\onecolumn

\appendix

\section{Glossary}
\label{sec:glossary}

Here, we define some terms that we use throughout the paper:
\begin{itemize}
\item \textbf{cisnormativity}: an ideology that promotes cisgender individuals as valid while painting transgender individuals as ``unnatural or dangerous'' \citep{nbwikiqueercishet}
\item \textbf{heteronormativity}: an ideology that promotes only heterosexual individuals as natural and normal \citep{10.1002/9781118896877.wbiehs205}
\item \textbf{hijra}: a third gender of ``male-bodied feminine-identified people'' \citep{Hossain02122017} rooted in the social, cultural and religious framework of India and several other South Asian countries \citep{hijradelhi}
\item \textbf{LGBTQIA+}: a collective term for gender, sex and sexuality minorities. It is an acronym for \textbf{L}esbians, \textbf{G}ay, \textbf{B}isexual, \textbf{T}ransgender, \textbf{Q}ueer/\textbf{Q}uestioning, \textbf{I}ntersexual and \textbf{A}romatic/\textbf{A}sexual, with the plus sign (\textbf{+}) including any other relevant minority identity not hereby covered
~\citep{HRCGlossary, nbwikiqueerlgbt}
\item \textbf{non-binary}: a category of individuals who do not fully identify with a binary gender i.e., individuals who are neither man or woman
\item \textbf{trans-exclusionary radical feminists (TERFS)}: individuals who advocate for ``feminism'' with the view that transgender women are infringing on the rights of cisgender women
\end{itemize}

\section{Language Breakdown}
\label{languages}
This is a breakdown of all the languages covered in the examined papers:



\begin{table}[H]
\begin{tabular}{lrll}
\toprule
Language of Data & Count & Singletons & \\
\midrule
English & 94 & code-mixed Urdu & Farsi \\
Italian & 20 & Korean & code-mixed English-Tamil \\
German & 15 & Telugu & Icelandic \\
Spanish & 13 & Kannada & Greek \\
French & 11 & Roman Urdu & Filipino \\
Portuguese & 5 & Maltese & Czech \\
Swedish & 5 & Indian English & Croatian \\
Chinese & 4 & Hindi & Turkish \\
Dutch & 4 & Catalan & Slovak \\
Danish & 4 & Polish & Gujarati \\
Slovene & 2 & Russian & \\
Tamil & 2 & Norwegian & \\
Norwegian & 2 &  & \\
\bottomrule
\end{tabular}
\caption{Counts of appearances of languages across papers. Languages with only one paper are grouped under ``Singletons.''}
 \label{tab:languages}
 \end{table}

\section{Limitations}
\label{sec:limitations}

In our efforts to survey the field of NLP for works regarding queer topics, we acknowledge two main limitations:
Firstly, while we include papers from other venues such as Interspeech, FAccT, CHI, etc.\@ in \cref{sec:common-themes} and non-English conferences in \cref{sec:non-english-queer-nlp}, our work is still centered on research published in *ACL conferences.
Secondly, our paper search utilized a list of queer identity keywords. 
Both factors might lead to a selection that misses publications from other venues, other non-English venues, and queer aspects that are not captured by our chosen keywords.

\section{Ethics Statement}\label{sec:ethics}

\paragraph{Positionality}
The authors of this paper hold many different standpoints afforded to us by our diverse identities, cultures, geographic locations, and personal histories. 
Most relevantly for this work, we draw from our positions both as experts in various NLP fields and as members and allies of queer communities. Our perspective on the issues we identify and discuss does not aspire to be an objective ``view from nowhere'' \citep{HarawayKnowledges}, but instead is rooted in our local knowledge and understandings.
In particular, this paper is largely shaped by North American and Eurocentric perspectives.
We draw together partial perspectives to synthesize a collective sense of the current state of Queer NLP, its implications and impacts, and possible futures. 

\section{Keywords}
\label{appendix:code}

\newcommand{\searchterm}[1]{\texttt{"#1"}}
We use the ACL Anthology Python library to retrieve all papers whose titles or abstracts contain at least one of the keywords below (case-insensitive matching with no lemmatization or word tokenization):
\searchterm{agender},
\searchterm{aromantic},
\searchterm{bisexual},
\searchterm{gay},
\searchterm{gender-ambiguous},
\searchterm{gender fair},
\searchterm{gender-fair},
\searchterm{gender inclusiv},
\searchterm{gender-inclusive},
\searchterm{gender-neutral},
\searchterm{gender non},
\searchterm{genderqueer},
\searchterm{glbt},
\searchterm{heteronormative},
\searchterm{homophobia},
\searchterm{homotransphob},
\searchterm{homotransphobia},
\searchterm{lesbian},
\searchterm{lgbt},
\searchterm{non-binary},
\searchterm{nonbinary},
\searchterm{queer},
\searchterm{same-gender},
\searchterm{trans-exclusionary}, 
\searchterm{transgender},
\searchterm{transphobia},
\searchterm{transphobic}.

\end{document}